\documentclass[12pt]{article}
\usepackage{epsfig,amsfonts,amssymb}
\begin{document}

\newcommand{\be}{\begin{equation}}
\newcommand{\ee}{\end{equation}}
\newcommand{\bea}{\begin{eqnarray}}
\newcommand{\eea}{\end{eqnarray}}

\baselineskip 14 pt

\begin{titlepage}
\begin{flushright}
{\small }
\end{flushright}

\begin{center}

\vspace{5mm}

{\Large \bf Mirage cosmology with an unstable probe D3-brane}

\vspace{3mm}

Dong Hyeok Jeong and Jin Young Kim \footnote{\tt
jykim@kunsan.ac.kr}

\vspace{2mm}

{\small \sl Department of Physics, Kunsan National University,
Kunsan, Chonbuk 573-701, Korea}

\date{\today}

\end{center}

\vskip 0.3 cm

\noindent

We consider the mirage cosmology by an unstable probe brane whose
action is represented by BDI action with tachyon. We study how the
presence of tachyon affects the evolution of the brane inflation.
At the early stage of the brane inflation, the tachyon kinetic
term can play an important role in curing the superluminal
expansion in mirage cosmology.

 \vskip 0.5 cm


\end{titlepage}

It is well known that the spectrum of string theory includes
unstable non-BPS D-branes besides the stable BPS D-branes. The
dynamics of unstable D-branes is described by the rolling of
tachyon \cite{sen}. The action is represented by Born-Infeld type
with decaying potential \cite{taction,bergshoeff,gskmm}. One can
study the time dependent solutions where tachyon rolls from the
top of the potential towards the minimum of the potential.
Recently Sen showed that classical decay of unstable D-brane
produces pressureless gas with non-zero energy density
\cite{sen2}. The possible effect of rolling tachyon to cosmology
was considered by Gibbons \cite{gibbons}. He took into account the
gravitational coupling by adding an Einstein-Hilbert term to the
effective action and showed that the cosmological evolution of an
accelerating universe can be obtained with the rolling of tachyon.
This model attracted attention in connection with various issues
of cosmology \cite{earlystcos}, despite the difficulty in the
simplest version of this theory \cite{KL}.

The brane world scenario opened a new frame to think about the
problems in cosmology \cite{addrs}. Many cosmological models based
on the brane universe have been studied widely in connection with
string theory \cite{kk,BDL,CRP,Keki,vsl,madridgp}. The idea of
brane universe is that our observed universe is a three-brane
(typically D-branes with maximal supersymmstry) embedded in a
higher dimensional space. The model of our interest is the mirage
cosmology suggested by Kehagias and Kiritsis \cite{Keki}. The key
idea of this model is that the motion of the brane through the
bulk, ignoring its back reaction to the ambient geometry, induces
cosmological evolution on the brane even when there is no matter
field on the brane. The crucial mechanism underlying the
construction of this formalism is the coupling of the probe brane
to the background fields.

This model was studied extensively in various directions
\cite{kimdilm,type0,youm}. In the previous work of one of the
authors and others \cite{type0}, the motion of a three-brane in
the background with tachyonic fields was studied in the framework
of type 0B string theory. It is shown that the presence of tachyon
in the background makes the effective matter density on the brane
less divergent compared with the case when there is no tachyon in
the background. In this letter we will consider the opposite case
that there is tachyon field in the probe brane while there is no
tachyon field in the background. The presence of tachyonic mode in
a probe brane indicates an instability of the probe brane. The
point of our interest is to observe how the presence of tachyon
affects the evolution of the brane inflation. As a concrete
example, we will consider the motion of a probe D3-brane in the
bulk background configuration of $\rm{AdS}_5 \times {\rm S}^5$.

We consider a system of supergravity fields coupled to a
(DBI+WZ)-type unstable brane which contains tachyon. To describe
the system in general, we need to consider the back reaction of
the brane to the bulk geometry. However we assume the probe brane
is light enough to neglect its back reaction. We consider the bulk
geometry whose action, keeping only the bosonic part, is
 \be
 S_{\rm bulk} = \frac{1}{2 \kappa^2} \int d^{10}x \sqrt{-g}
 \left[ e^{-2\phi} (R +4 (\nabla \phi)^2 ) - \frac{1}
 {2(n+1)!} F_{n+1}^2 \right],  \label{bulkaction}
 \ee
where $\phi$ is the dilaton field and $F_{n+1}$ is the field
strength tensor of RR potential $C_n$ made by a stack of many
stable D-branes. We will present our analysis with $n=4$ case
where the background geometry is ${\rm AdS_5 \times S^5}$ form.
The metric can be parameterized as
 \be
 ds^2 = g_{00}(r) dt^2 + g(r) d {\vec x}^2 + g_{rr}(r) dr^2 + g_S(r)
 d\Omega_5^2 ,
 \ee
where $(t, \vec x )$ are the three brane coordinates and $r$ is
the radial coordinate perpendicular to the brane.

Unstable Dp-branes are described by rolling tachyon. Coupling
tachyon effective action to gravity has been considered by many
authors \cite{bergshoeff,coupltg,bucwal}. We start from the action
of \cite{bucwal} described by a (DBI+WZ)-type
 \be
 S_{\rm p-brane} = - T_{p} \int d^{p+1} \xi
 \left [ e^{-\phi} V(T) \sqrt{-{\rm det} ({\hat A}_{\mu\nu}) }
 + f(T) dT \wedge {\hat C}_p \right],   \label{Spbrane}
 \ee
where
 \bea
 {\hat A}_{\mu\nu} &=& A_{MN} \frac {\partial x^M}{\partial x^\mu}
 \frac{\partial x^N}{\partial x^\nu} , \nonumber  \\
 A_{MN} &=& g_{MN} + \partial_M T \partial_N T .
 \nonumber
 \eea
Here $M,N$ are the 10-dimensional bulk indices and $\mu,\nu$ are
the brane ones. We neglect all fields other than gravity and
tachyon for simplicity. The forms of coupling $V(T)$ and $f(T)$
are not known precisely. Since the tachyon potential $V(T)$
measures the varying tension, it should satisfy two boundary
values such that $V(T=0)=1$ and $V(T=\infty)=0$. Specific
computation of string theory \cite{gskmm} and a plausible argument
based on it \cite{sen} predict $V(T)\sim e^{-T^{2}}$ for small $T$
and $V(T)\sim e^{-T}$ for large $T$. The potential $V(T)$ is a
smooth function in the intermediate region and connects two
asymptotic expressions.Thus the following shape of the potential
is widely adopted in tachyonic cosmology for the sake of
simplicity
\begin{equation}\label{V3}
V(T)=\frac{V_0 }{{\rm cosh}\left(\frac{T}{T_0}\right)},
\end{equation}
where $T_{0}$ is determined by string theory. We will focus on the
early stage of the universe where the precise forms of $V(T)$ and
$f(T)$ do not affect the argument significantly.

The WZ term of the brane action contributes only when there is
three-from potential in the bulk. Since we consider the case when
the probe brane is three-dimensional and there is only four-form
potential in the bulk which makes the geometry of ${\rm AdS_5
\times S^5}$, this term is neglected. Thus the three brane action
can be simplified as
 \be
 S_3 = - T_{3} \int d^{4} \xi
  e^{-\phi} V(T) \sqrt{-{\rm det} ({\hat A}_{\mu\nu}) }.
 \label{s3bract}
 \ee
The above brane action is a good approximation both for the
perturbative vacuum region ($T=0$) and stable vacuum region
($T=\infty$). The rolling tachyon solution describing the unstable
D-branes is time reversal symmetric. So one can take the initial
condition $\dot T (t=0)=0$ without loss of generality. Since $\dot
T=0$ initially, there can exist a time interval where $T$ remains
fixed near its perturbative vacuum ($T=0$). We confine our
attention on this stage of the cosmological evolution.

To write down the action explicitly, we take the worldvolume
coordinates $\xi^\alpha = x^\alpha ~ (\alpha = 0,1,2,3)$ in the
static gauge. We assume that the tachyon field $T$ does not have
spatial dependence $T =T(t)$. It has no $r$ dependence since
tachyon exists only on the probe brane. In general, the motion of
a probe D3-brane can have a nonzero angular momentum in the
transverse directions. Then the Lagrangian of the brane can be
expressed as
 \be
 L = -T_3 V(T) e^{-\phi} \sqrt{ g^3 ( |g_{00}(r)| - {\dot T}^2
 -g_{rr} {\dot r}^2 - g_S h_{ij} {\dot \varphi}^i
 {\dot \varphi}^j ) } ,
 \label{lag1}
 \ee
where the dot denotes the derivative with respect to $t$ and $
h_{ij} \varphi^i \varphi^j $ is the line element on the unit five
sphere ($i,j=5, \dots, 9$). For convenience of notation, we
parameterize the Lagrangian, divided by three brane tension $T_3$,
as
 \be
 L = - \sqrt{ A(r) - K(r) {\dot T}^2 - B(r) {\dot r}^2
 - D(r) h_{ij} {\dot \varphi}^i {\dot \varphi}^j } ,
 \label{lag2}
 \ee
where
 \bea
 A(r) &=& V^2(T) g^3(r)|g_{00}(r)|e^{-2\phi}, \nonumber  \\
 K(r) &=& V^2(T) g^3(r)e^{-2\phi},  \nonumber  \\
 B(r) &=& V^2(T) g^3(r)g_{rr}(r)e^{-2\phi},  \nonumber  \\
 D(r) &=& V^2(T) g^3(r)g_S(r)e^{-2\phi}.  \nonumber
 \label{ABD}
 \eea

 The momenta of the system are given by
 \bea
 && p_r = {B(r) \dot r \over
 \sqrt{A(r)-K(r) {\dot T}^2- B(r)\dot r^2 - D(r) h_{ij} {\dot \varphi}^i
 {\dot \varphi}^j } } ,
    \nonumber \\
 && p_i  = {D(r) h_{ij} \dot \varphi^j \over
 \sqrt{A(r)-K(r) {\dot T}^2- B(r)\dot r^2 - D(r) h_{ij} {\dot \varphi}^i
 {\dot \varphi}^j } } .
 \eea
Calculating the Hamiltonian and demanding the conservation of
energy, we have
 \be
 H =  { A(r) \over \sqrt{A(r)- K(r) {\dot T}^2- B(r)\dot r^2-D(r)h_{ij} {\dot
 \varphi}^i {\dot \varphi}^j} } = E ,
 \label{Hamil}
 \ee
 where $E$ is the total energy of the brane.
 Also from the conservation of the total angular momentum $h^{ij}p_ip_j=\ell^2$,
 we have
 \be
 h_{ij} {\dot \varphi}^i {\dot \varphi}^j =
 {\ell^2[A(r)-K(r) {\dot T}^2-B(r)\dot
 r^2] \over  D(r)[D(r)+\ell^2]} .
 \label{lcons}
 \ee
The equation of motion for the tachyon field is
 \bea
 && \partial_t \left(
 \frac {K{\dot T}}{\sqrt{A(r)- K(r) {\dot T}^2- B(r)\dot r^2-D(r)h_{ij} {\dot
 \varphi}^i {\dot \varphi}^j} } \right)  \nonumber \\
 && - \frac{V V^\prime g^3 e^{-2 \phi} {\dot T}^2 }
  {\sqrt{A(r)- K(r) {\dot T}^2- B(r)\dot r^2-D(r)h_{ij} {\dot
 \varphi}^i {\dot \varphi}^j} } = 0,  \label{taceom}
 \eea
where $V^\prime = {dV \over dT}$.

We can consider another form of WZ term $-T_3 \int d^{4} \xi \hat
C_{4}$ which does not contain tachyon. However it is known that
the presence of this term results in the shift of the total energy
of the brane \cite{Keki}. Adding this four-form potential in the
brane action (\ref{Spbrane})

 \be
 S_3 = - T_{3} \int d^{4} \xi [
  e^{-\phi} V(T) \sqrt{-{\rm det} ({\hat A}_{\mu\nu})} + \hat
C_{4} ],
 \ee
 and defining $C(r) = C_{0123} (r)$, the Lagrangian
(\ref{lag2}) is modified to

 \be
   L = - \sqrt{ A(r) - K(r) {\dot T}^2 - B(r)
{\dot r}^2
 - D(r) h_{ij} {\dot \varphi}^i {\dot \varphi}^j } + C(r).
 \ee
 From this, the Hamiltonian is calculated as

\be
 H = { A(r) \over \sqrt{A(r)- K(r) {\dot T}^2- B(r)\dot
r^2-D(r)h_{ij} {\dot \varphi}^i {\dot \varphi}^j} } - C.
 \ee
 If we require the conservation of energy, the modification is just the
 shift of energy $E \to E + C$. It will be straightforward to
 extend the above result when we turn
on the NS/NS field $B_{\mu\nu}$ and gauge field $F_{\mu\nu}$ on
the probe brane.

In principle if we know the potential we can solve the coupled
system of equations (\ref{Hamil}), (\ref{lcons}) and
(\ref{taceom}), and use this result to proceed further. However,
when the tachyon is rolling from the top of the potential where $
{dV \over dT} = 0$, we can neglect the second term as a first
approximation. Then the role of tachyon on the probe brane is
similar to that of a scalar field. In this case we have
 \be
 \frac {K{\dot T}}{\sqrt{A(r)- K(r) {\dot T}^2- B(r)\dot r^2-D(r)h_{ij} {\dot
 \varphi}^i {\dot \varphi}^j} } \simeq Q,  \label{tacchgcons}
 \ee
where $Q$ is an integration constant. By solving Eqs
(\ref{Hamil}), (\ref{lcons}) and (\ref{tacchgcons}) in terms of
${\dot T}^2$, ${\dot r}^2$ and $h_{ij} {\dot \varphi}^i {\dot
\varphi}^j$, we have
 \bea
 {\dot T}^2 &=& \frac{Q^2}{K^2} \frac{A^2}{E^2}  ,
 \label{tdotsquare}   \\
 {\dot r}^2 &=& \frac{A}{B} \left( 1 - \frac{KD + K\ell^2 + Q^2
 D}{KD}
 \frac{A}{E^2} \right) , \label{rdotsquare}  \\
 h_{ij} {\dot \varphi}^i {\dot \varphi}^j &=& \frac{\ell^2}{D^2}
 \frac{A^2}{E^2} .  \label{angdotsquare}
 \eea

In the point of an observer living on the brane, the induced
metric on the brane is the natural metric to see the evolution of
its brane. The induced metric on the probe 3-brane universe is
 \be
 d \hat s^2_{4d} = (g_{00}+g_{rr} \dot r^2
 + g_S h_{ij} {\dot \varphi}^i {\dot \varphi}^j )
 dt^2 + g(d\vec x)^2 .
 \label{4dmet}
 \ee
Substituting Eqs. (\ref{rdotsquare}) and (\ref{angdotsquare}), the
induced metric is calculated as
 \be
 d\hat s^2_{4d}= - \frac {g^2_{00} ( g^3e^{-2\phi}V_0^2 + Q^2 )} {E^2} dt^2
 + g (d\vec x)^2 .
 \ee
We define, for the standard form of a flat expanding universe, the
cosmic time $\eta$ as
 \be
 d \eta= \frac {|g_{00}| \sqrt{ g^3e^{-2\phi}V_0^2 + Q^2 } } {E} dt.
 \ee
If we define the scale factor as $a^2=g$, we can calculate from
the analogue of the four-dimensional Friedman equation the Hubble
constant $H={\dot a / a}$
 \be
 \left({\dot a \over a}\right)^2
 = \frac{E^2 - ( g^3e^{-2\phi}V_0^2 + Q^2 + \ell^2 g_S^{-1} ) |g_{00}|}
 {4|g_{00}| g_{rr} ( g^3e^{-2\phi}V_0^2 + Q^2 )}
 \left({g'\over g}\right)^2 ,
 \label{hub}
 \ee
 where the dot denotes the derivative
with respect to cosmic time $\eta$ and the prime denotes the
derivative with respect to $r$. The right hand side of Eq.
(\ref{hub}) can be interpreted as the effective matter density on
the probe brane
 \be
 {8\pi \over 3}\rho_{\rm eff} =
 \frac{E^2 - ( g^3e^{-2\phi}V_0^2 + Q^2 + \ell^2 g_S^{-1} ) |g_{00}|}
 {4|g_{00}| g_{rr} ( g^3e^{-2\phi}V_0^2 + Q^2 )}
 \left({g'\over g}\right)^2 .
 \label{hubble}
 \ee
The key idea of the above mechanism can be summarized as follow.
Though the tachyon does not appear explicitly in the induced
metric, it affects the cosmic time through the coupling with $r$
and $\varphi$ equation.

As an application of the above formalism, we will consider the
cosmology of the probe unstable D3-brane under the background
configuration given by (\ref{bulkaction}). The near-horizon
geometry, corresponding to black hole solution, is ${\rm AdS_5
\times S^5}$ form with the metric
 \be
 ds^2 = {r^2 \over L^2} ( -f(r) dt^2 + (d\vec x )^2 )
  + {L^2 \over r^2} \frac{dr^2}{f(r)} + L^2 d\Omega_5^2,
 \ee
 where $f(r) = 1 - ( r_0 / r )^4$ and $r_0$ is the location of the horizon.
 The effective density on the probe D3-brane is calculated as
 \be
 {8\pi \over 3}\rho_{\rm eff} =
 \frac{1}{ a^2 ( a^6 e^{-2\phi}V_0^2 + Q^2 ) L^2}
 \left[ E^2 -
 \left ( 1 -  \frac{r_0^4}{L^4} \frac{1}{a^4} \right)
 a^2 ( a^6e^{-2\phi}V_0^2 + Q^2 + {\ell^2 \over L^2} )
 \right].
 \label{effdensity}
 \ee

In the limit $Q=0$, $V_0 = 1$ and $\phi=0$, which corresponds to
mirage cosmology with stable (nontachyonic) probe brane under the
nondilatonic background, the above result exactly recovers the
known result (see Eq. (5.2) of \cite{Keki} with $C=0$)
 \be
 {8\pi \over 3}\rho_{\rm eff} =
 \frac{1}{ L^2}
 \left[ \frac{E^2}{a^8} -
 \left ( 1 -  \frac{r_0^4}{L^4} \frac{1}{a^4} \right)
 \left( 1 + {\ell^2 \over L^2} {1 \over a^6 }\right)
 \right].
 \label{effdennontac}
 \ee
For the nontachyonic probe brane, the power $a^{-8}$ dominates at
earlier time. This corresponds to the superluminal equation of
state parameter $\omega = 5/3$ ($p = \omega \rho$). Note that this
power is smoothed by the tachyonic charge $Q$ in
(\ref{effdensity}).

The presence of tachyon on the probe brane makes the expansion
less divergent as $a \to 0$ where our assumptions are valid. In
general, if we add any matter field, the brane universe will
inflate faster due to the increased effective density. However, in
\cite{type0}, it is shown that the presence of {\sl tachyonic
field in the bulk background} makes the expansion of the brane
less divergent compared with the case without tachyon. Our
analysis shows that the presence of {\sl tachyonic field in the
probe brane} can also make the expansion less divergent. So we
conclude that the presence of tachyon field which carries the
instability of the brane, whether it exists in the bulk background
or in the probe brane, makes the mirage inflation less divergent.

In the point of higher dimensional theory, our approximation is
valid near the horizon of the higher dimensional black hole made
by a stack of heavy stable branes. Conditions in the higher
dimensional theory are translated as the constraints in the
parameter space. For example, since ${\dot r}^2 \ge 0$ for
equation (\ref{rdotsquare}) this gives the constraint
 \be
 \frac{A}{B} \left( 1 - \frac{KD + K\ell^2 + Q^2
 D}{KD}  \frac{A}{E^2} \right) \ge 0 .
 \ee
Another example is the brane Hamiltonian. For the simple case we
considered in this paper, the positivity of energy is
automatically satisfied (see Equation(\ref{Hamil})). When we add a
four-form potential as mentioned before, the allowed values of $r$
is determined by the energy constraint $C(r) + E \ge 0 $. Similar
argument holds when we consider $B_{\mu\nu}$ and $F_{\mu\nu}$.

We would like to point out that our result is valid only when the
tachyon is near the top of the potential so that the approximation
(\ref{tacchgcons}) can be applied. At the early stage of mirage
cosmology as we considered here, the effect of tachyon potential
is neglected. Then the tachyon kinetic term plays the dominant
role for the brane inflation. At late time where one recovers the
usual vacuum gravity, it is expected that the cosmology will be
driven not only by the brane energy but also by the bulk energy.
The behavior of the cosmology during the intermediate time can be
studied by solving the tachyon equation of motion (\ref{taceom})
together with (\ref{Hamil}) and (\ref{lcons}). In this region both
the kinetic term and potential slope should be considered.
However, finding the analytic solution seems very difficult for
the given potential form. Numerical solution may give some hints.

The intuitive reasoning of the scalar modification to the FRW
equation is similar to the modification by a gauge electric field.
As far as the tachyon potential is constant, the effect of tachyon
is a scalar field affecting the equation of motion of the brane
evolution. This case is similar to mirage cosmology with an
electric field on the probe brane. When there is an electric
field, the square root of BDI action is given by (Equation (4.1)
of Ref. \cite{Keki})

 \be
 -\det({\hat G}_{\mu\nu} + 2 \pi \alpha^\prime{\hat
F}_{\mu\nu}) = g^3 ( |g_{00}(r)| - g_{rr} {\dot r}^2 - g_S h_{ij}
{\dot \varphi}^i {\dot \varphi}^j - {\varepsilon}^2 g^2 ),
 \ee
 where ${\varepsilon}^2 = (2 \pi \alpha^\prime)^2 E_i E^i$.
Comparing the square root of Equation (\ref{lag1}) with the above
equation, the only difference is that there is no metric coupling
($g^2$ factor) in tachyon kinetic term. Thus the mechanism how the
tachyonic nature of the scalar enters into the final FRW-like
equation is the same as the case with U(1) electric field on the
probe brane. The integration constant $\mu^2 = (2 \pi
\alpha^\prime)^2 \mu_i \mu^i$, obtained from the equation of
motion for the electric field, enters the final FRW-like equation
(Equation(4.7) of Ref.\cite{Keki}). Similarly the tachyonic nature
of the scalar is carried by $Q^2$ (Equation(\ref{effdensity})).

It has been shown that, for large $T$, the value of the tachyon
field at a given space-time point can be identified as the time
coordinate when tachyon is coupled only to gravity
\cite{sentactime}. Further studies on mirage effect for large
value of tachyon field are expected.

\vspace{.5cm} {\bf Acknowledgements} This work was supported by
the Korea Reserch Foundation Grant (KRF-2004-002-C00065).  J.Y.
Kim is grateful to the Department of Physics at U.C. Davis for
hospitality during his visit.

\end{document}